\documentclass[12pt,showpacs,amsmath,amssymb,prd,superscriptaddress]{revtex4}
\usepackage{epsf}
\usepackage[dvips]{graphicx}
\usepackage[normalem]{ulem}

\begin{document}

\title{
$1/f$ fluctuations in spinning-particle motions around a Schwarzschild black hole
}

\date{\today}

\author{Hiroko Koyama\footnote{koyama@gravity.phys.waseda.ac.jp}}
\affiliation{Department of Physics,
Waseda University, Shinjuku-ku, Tokyo, 169-8555, Japan}
\affiliation{Department of Physics, Nagoya University, Nagoya, 464-8602, Japan}
\author{Kenta Kiuchi\footnote{kiuchi@gravity.phys.waseda.ac.jp}}
\affiliation{Department of Physics,
Waseda University, Shinjuku-ku, Tokyo, 169-8555, Japan}
\author{Tetsuro Konishi\footnote{tkonishi@r.phys.nagoya-u.ac.jp}}
\affiliation{Department of Physics, Nagoya University, Nagoya, 464-8602, Japan}

\begin{abstract}
We study the properties of chaos in the motions of a spinning test
 particle in Schwarzschild spacetime.
We characterize the chaos using the power spectrum of the time series of
 $z$ components of the particle's position.
It is found that the pattern of 
the power spectrum shows not only white noise but also $1/f$-type
 fluctuation,
depending on 
the value of the total angular momentum $J$ and 
the spin $S$ of the test particle.
Therefore
we succeed in classifying the chaotic motions, which have been
classified as simply chaotic ones in former works, into the two distinct
 types.
One is $1/f$, and the other is white noise.
Based on this classification, we plot,
in the two-dimensional parameter space $(J,S)$,
the phase diagram for the properties of the chaos.
This phase diagram enables us in principle to guess the properties of the
 system $(J,S)$ by observing the dynamics of the test particle, even if
 the motion is chaotic.
Furthermore, 
we detect that the origin of the $1/f$ fluctuation is
that the particle motion stagnates around regular orbits (tori), while
 traveling back and forth between them, which is called 
``stagnant motion'' or ``sticky motion'' in Hamiltonian dynamical systems.
The point is that
the difference of the property of the chaos or the power spectra
is due to the topological structure of the phase space, which in
 turn is governed by the physical parameter set $(J,S)$ of the system.
From this point of view, the chaos we found in this system is not always
 merely random.
\end{abstract}


\maketitle

\section{Introduction}
Nature is filled with phenomena that exhibit chaotic behavior.
In chaotic systems we cannot predict the system's future state exactly~\cite{ll92,co02}.
Such chaotic behavior has also been found in some relativistic systems 
~\cite{Hobill,Barrow,Conto,Dettmann,Yust,Karas,Varvoglis,Bombelli,Moeckel,Sota,
sm97,Suzuki2,Suzuki3,Mich,Seme,Levin,Schnittman,Cornish,km04}.
For example, in Schwarzschild spacetime, the motions of a spinning test 
particle can be chaotic~\cite{sm97}.
If the test particle does not have spin, the motion of the test particle is regular.
They have found that, as the magnitude of the spin increases
extremely, 
the motions switch from regular to chaotic, using the Poincar\'{e} map
and the Lyapunov exponent.
In practice, the magnitude of the spin where the particle motion is remarkably
chaotic is so large that such a system is not realistic, 
which has been remarked in the paper~\cite{sm97}. 
However, this model is important in understanding chaos in general
relativistic systems.

In this paper, we
look for statistical laws to characterize the chaos
in the motions of the spinning
test particle in Schwarzschild spacetime.
Actually, the chaotic motions in this system have been classified merely
as chaotic, according
to the distribution of the points in Poincar\'{e} maps and positiveness
of the Lyapunov
 exponents, but the details of the properties of the chaos
have not been clarified~\cite{sm97}.
Once we find chaotic behavior, however, 
we should rather characterize it to extract the specific properties
of the system.
Indeed, we can hardly learn anything about the chaos if we judge it only
from the randomness of the distribution of the points in Poincar\'{e} maps 
or the positiveness of the Lyapunov exponents~\cite{ll92,co02,Hobill}.
Not a few people believe that a chaotic system is simply random and 
completely unpredictable.
It is true that we cannot predict 
the time evolution of the state of the system exactly, 
when the system is chaotic.
However, we should note that,
even in such cases, we can frequently find some statistical laws 
which are specific to the system. 
One possible measure of chaos is the power spectrum of the time series
of the system.
If the power spectrum is white noise,
the time evolution of the system is not time-correlated.
In not a few cases, however,
the pattern of the power spectrum obeys power laws, so-called $1/f$
fluctuations~\cite{dh81,weissman88,schuster84,manneville80,kohyama84,geisel87}, 
which can be clearly distinguished from the white-noise type.
That the power spectrum obeys some power laws means 
that the time evolution of the system is time-correlated.
Thus we can classify the chaos from the pattern of its power spectrum.

Let us explain our strategy to characterize the chaos 
in the motions of the spinning test particle in Schwarzschild spacetime
in this paper. 
To begin with, we introduce the power spectrum of the time series of $z$ components
of the particle's position.
Next we characterize the properties of the chaos,
using the pattern of the power spectrum.
It is found that the pattern of the power spectrum can be classified as
$1/f$ or white noise.
That is,
we succeed in classifying the chaotic motions into two distinct types.
Furthermore, based on this classification,
we can plot, in the two-dimensional parameter space $(J,S)$,
the phase diagram for the properties of the chaos.
Finally, we detect
the origin of the $1/f$ fluctuations of the power spectrum in this
system. We find out that the orbit stagnates around the tori,
while traveling back and forth between them,
whenever the power spectrum shows $1/f$ spectral pattern.

This paper is organized as follows.
In Sec.~\ref{sec:model} we shall briefly review the basic equations, i.e., the equations
of motion for a spinning test particle in Schwarzschild spacetime.
In Sec.~\ref{sec:poin} we plot the Poincar\'{e} maps of the chaotic
motions in this system. 
Here we point out a weakness in the Poincar\'{e} maps, which
brings a motivation to introduce another method
to characterize such chaotic motions.
In Sec.~\ref{sec:ps} we introduce the power spectrum
to characterize the properties of the chaos in this system.
Then we find the pattern of the power spectrum can be classified into two
types, $1/f$ and white noise, depending on the values of the system parameters.
In Sec.~\ref{sec:origin} we detect the origin of the $1/f$ fluctuations.
The final section is devoted to summary and discussion.
Throughout this paper we use units $c=G=1$

\section{Equations for a spinning test particle in Schwarzschild spacetime}
\label{sec:model}
We consider a spinning test particle in Schwarzschild spacetime,
\begin{equation}
ds^2=-\left(1-\frac{2M}{r}\right)dt^2+\left(1-\frac{2M}{r}\right)^{-1}
dr^2
+r^2d\theta^2+r^2\sin^2\theta d\phi^2,
\end{equation}
where $M$ is the mass of the black hole.
The equations of motions of a spinning test particle in relativistic
spacetime have been derived by Papapetrou \cite{papa} and then reformulated
by Dixon \cite{dixon}.
The set of equations is given as
\begin{equation}
 \frac{dx^\mu}{d\tau} = v^\mu, \label{basic1}
\end{equation}
\begin{equation}
\frac{Dp^\mu}{d\tau} = - \frac{1}{2}{R^\mu}_{\nu\rho\sigma}  v^\nu
S^{\rho\sigma}, \label{basic2}
\end{equation}
\begin{equation}
\frac{D S^{\mu\nu}}{d\tau} = 2 p^{[\mu}
v^{\nu]}, \label{basic3}
\end{equation}
where $\tau,v^\mu,p^\mu$, and $S^{\mu\nu}$ are an affine parameter  of the
orbit, the four-velocity of a particle, the momentum, and the spin
tensor, respectively. $p^\mu$ deviates from a geodesic due to the
coupling of the Riemann tensor with the spin tensor.
We adopt the additional condition formulated by Dixon~\cite{dixon},
\begin{equation}
p_\mu S^{\mu\nu} = 0 \label{cond1},
\end {equation}
which gives a relation between $p^\mu$ and $v^\mu$, 
and consistently determines the center of mass of the spinning particle.
The mass of the particle $\mu$ is defined by 
\begin{equation}
\mu^2 = - p_\mu p^\mu.\label{const1}
\end{equation}
To make clear the freedom of this 
system, we have to check the conserved quantities. Regardless of the
symmetry of the background spacetime, it is easy to show that the mass $\mu$ and
the magnitude of spin $S$ defined by
\begin{equation}
S^2 \equiv \frac{1}{2}S_{\mu\nu} S^{\mu\nu}
\end{equation}
are constants of motion \cite{wald72}. If a background spacetime possesses some
symmetry described by a Killing vector $\xi^\mu$,
\begin{equation}
\label{eqn:killing}
C_\xi \equiv \xi^\mu p_\mu - \frac{1}{2}\xi_{\mu;\nu} S^{\mu\nu}
\end{equation}
is also conserved~\cite{dixon}.  
Because the spacetime we consider in this paper is static
and spherically symmetric, there are  two Killing vector fields,
$\xi_{(t)}^{\mu}$ and
$\xi_{(\phi)}^{\mu}$. From (\ref{eqn:killing}), we find the
constants of motion related with those Killing vectors as 
\begin{eqnarray}
E & \equiv & - C_{(t)}=-p_t-\frac{M}{r^2}S^{tr},
\label{eqn:energy}
\\
J_{z} & \equiv & C_{(\phi)}=p_{\phi}-r (S^{\phi
r}-rS^{\theta\phi}\cot\theta )\sin^2\theta.
\label{eqn:jz}
\end{eqnarray}
$E$ and $J_z$ are interpreted as the energy of the particle and
the $z$ component of the total angular momentum,  respectively.
Because the spacetime is spherically symmetric, the  $x$ and $y$
components of the total angular momentum are also
conserved. 
In addition, without loss of generality
we can choose the $z$ axis in the direction of total angular momentum as 
\begin{equation}
(J_{x},J_{y},J_{z})=(0,0,J),
\label{eqn:total}
\end{equation}
where $J>0$. In the following sections,
we integrate the above equations of motion
numerically for various values of parameters $E$, $J$, and $S$,
using the Bulirsch-Stoer methods \cite{bs}.

\section{Poincar\'{e} maps}
\label{sec:poin}
In this section, we illustrate the particle motions 
in the model formulated in the previous section
(see also \cite{papa,dixon,sm97}) using Poincar\'{e} maps 
as shown in Fig.~\ref{fig:poin}.
As mentioned in the paper \cite{sm97},
the parameter range of the energy $E$ for each fixed parameter set $(J,S)$
is restricted to enable the  particle to move around the black hole
without going to infinity or falling into the black hole.
Although similar analyses have been already done in the paper \cite{sm97}, 
the analysis in this section motivates us to introduce in the next section
another method to characterize the motions.

Using Poincar\'{e} map, we can judge if the motions are chaotic or not.
To plot the Poincar\'{e} map, we adopt the
equatorial plane ($\theta =
\pi/2$) as a Poincar\'{e} section and plot the point ($r$, $v^r$)  when the
particle crosses the Poincar\'{e} section with $v^\theta < 0$ (Fig.~\ref{fig:poin}). 
In Fig.~\ref{fig:pmap}, we plot the Poincar\'{e} maps 
for the total angular momentum $J=4\mu M$.
The values of the spin $S$ are set to $S=1.2\mu M$ in Fig.~\ref{fig:pmap} (a) and
$S=1.4\mu M$ in Fig.~\ref{fig:pmap} (b).
Each value of the energy $E$ is chosen as an appropriate 
one so that the test particle does not escape to infinity and does not
fall into a black hole.
The parameter sets $(J,S,E)$ in Figs.~\ref{fig:pmap} (a) and
\ref{fig:pmap} (b) 
are the same as those in Figs. 4 (e) and 4 (f) in the paper~\cite{sm97}.
In Fig.~\ref{fig:pmap} dots with different colors correspond to the data from the orbits with
different initial conditions.
If the orbit is chaotic, some of the tori are broken and the
Poincar\'{e} map no longer consists of a set of closed curves.
Both in Fig.~\ref{fig:pmap} (a) and Fig.~\ref{fig:pmap} (b), 
the points in the Poincar\'{e} maps are scattered randomly,
and the so-called chaotic sea is formed. 
That is, the orbits with both sets of parameters are chaotic.

Now it is worthwhile to note a weakness in the Poincar\'{e} map.
The chaotic sea in Figs.~\ref{fig:pmap} (a) and \ref{fig:pmap} (b) 
cannot be distinguished apparently.
Indeed, the existence of chaotic sea in the Poincar\'{e} map 
allows us to judge whether the particle motions in this model
are regular or chaotic~\cite{sm97}.
The positiveness of the Lyapunov exponent, which for this model
have been also investigated in~\cite{sm97}, does too.
That is, we can tell that the motion is chaotic using these measures.
However, only from these measures,
we cannot know more than that the motion is merely chaotic,
when it is chaotic.
Once we find the chaotic behaviors, we should rather characterize the
chaos, since being chaotic does not always mean randomness or no rule.
Therefore, another method is necessary to characterize the chaos
in more detail.
In the next section, we will introduce the power spectrum
to classify such chaotic motions.

\section{$1/f$ fluctuations of the power spectrum}
\label{sec:ps}
In this section, we characterize the chaos in the spinning test particle motions
which was shown in the previous section.
Here we analyze the time series of the particle position.
In order to do that, first of all, we introduce the power spectrum.
The power spectrum of the time series of $z$ components of the
particle's position,
$P_z(\omega)$, is defined by
\begin{equation}
 P_z(\omega)\equiv \left|\int_{0}^{T} z(t)e^{i\omega t}dt
\right|^2.
\end{equation}
Here we set $T=10^5$ in our computation.
If we define the autocorrelation function $\Phi_z(\tau)$ as
\begin{equation}
\label{eq:autocor}
\Phi_z(\tau)\equiv \frac{1}{2T}\int_{-T}^{T} z(t)z(t+\tau)dt,
\end{equation}
and take the limit $T\to \infty$,
Wiener-Khinchin's theorem~\cite{leichel} relates the power spectrum
$P_z(\omega)$ and the autocorrelation function $\Phi_z(\tau)$ as
\begin{equation}
\label{eq:wh}
 P_z(\omega)=\int_{-\infty}^{\infty} \Phi_z(\tau)e^{-i\omega \tau}d\tau.
\end{equation}

We plot the power spectrum $P_z(\omega)$ in Fig.~\ref{fig:ps}.
In Figs.~\ref{fig:ps} (a) and \ref{fig:ps} (b), 
we choose parameter sets with the same values as those
in Figs.~\ref{fig:pmap} (a) and \ref{fig:pmap} (b), respectively.
Each line color in Figs. \ref{fig:ps} (a) and \ref{fig:ps} (b)
also corresponds to that of the dots in Figs.~\ref{fig:pmap} (a) and
 \ref{fig:pmap} (b), respectively.
Note that the patterns of the power spectra are
different between Fig.~\ref{fig:ps} (a) and Fig.~\ref{fig:ps} (b).
In particular, we find that one is $1/f^\nu$ $(f\equiv \omega/2\pi)$ where $\nu\simeq
1.2$ (Fig.~\ref{fig:ps}(a)), 
while the other is white noise (Fig.~\ref{fig:ps}(b)) 
in the low-frequency range about $\omega \leq 0.01$. 
Using Eq. (\ref{eq:wh}), we see that
if the power spectrum is of $1/f^\nu $ type, 
the temporal correlation is also of the power-law type,
which means a strong temporal correlation with no typical time scale.
On the other hand, if the power spectrum is of white noise type, temporal correlation
is a $\delta$-function, which means no temporal correlation.
Therefore we can clearly distinguish the chaotic motions by using the pattern of the power
spectrum.
Moreover, Fig~\ref{fig:ps} shows that
the pattern of the power spectrum $P_z(\omega)$,
that is the property of the chaos,
depends on the system parameters, 
the total angular momentum $J$ and the spin $S$ of the test particle.
In addition, as shown in Fig~\ref{fig:ps},
the pattern of the power spectrum is independent of the initial
conditions, if we fix the parameter set.

Furthermore, we test the pattern of the power spectrum $P_z(\omega)$ for various
grid points in the two-dimensional space $(J,S)$, and obtain a
new phase diagram
for the properties of the chaos, summarized in Fig.~\ref{fig:pdia}.
Each value of the energy $E$ is chosen as an appropriate 
one so that the test particle does not escape to infinity and does not
fall into the black hole, and the chaotic motions
that we pay attention to in this paper are dominant.
In the blank region in Fig.~\ref{fig:pdia},
the test particle goes to infinity or falls into a black hole,
since the energy surface is unbounded.
This blank region agrees with the region of the type (U2) in Fig.3 
in the paper~\cite{sm97}.
In the region where circle ($\bigcirc$) and cross ($\times$) symbols 
are marked in Fig.~\ref{fig:pdia}, 
the particle motion is chaotic and bounded.
In the region where circle symbols ($\bigcirc$) are marked
in Fig.~\ref{fig:pdia}, the power spectrum $P_z(\omega)$
shows $1/f$-type spectrum. On the other hand, 
in the region where cross symbols ($\times$) are marked,
the power spectrum $P_z(\omega)$ shows white noise.
The total region together with dotted and crossed regions
is included in the region of the type (B2) in Fig.3 in the paper \cite{sm97}.
In the paper \cite{sm97}, the motions in the region of the type (B2)
are classified as to be merely chaotic ones.
Therefore, 
Fig.~\ref{fig:pdia} means that
we succeed in classifying the  chaotic motions for various values of parameter
set $(J,S)$, which have been classified as simply chaotic in former works,
into two distinct types, $1/f$ and white noise, using the power spectrum $P_z(\omega)$.

Now we consider the physical meanings of the tendency of
the power spectrum pattern to
change from $1/f$ to white noise by increasing the
magnitude of spin $S$ of the test particle in Fig.~\ref{fig:pdia}.
Let us note that 
the system is integrable and the motion of the test particle is not chaotic,
if the test particle does not have spin ($S=0$) \cite{co02,sm97}.
Furthermore, the motion of the test particle with the small magnitude of
spin remains almost regular~\cite{KAM-K,KAM-A,KAM-M}.
The motion become remarkably chaotic at last, if the magnitude of the
spin is extremely large~\cite{sm97}.
Then it is reasonable to regard
the magnitude of the spin $S$ of the test particle as a measure of chaos.
Considering these facts,
our results can be interpreted to mean that 
the pattern of the power spectrum $P_z(\omega)$
changes from $1/f$ type to white noise 
as the strength of the chaos increases.
Therefore, Fig.~\ref{fig:pdia} means that
we succeed in classifying the chaotic motions into two categories
in accordance with the strength of chaos.
In the next section, we investigate the origin of the $1/f$
fluctuations in the power spectrum $P_z(\omega)$.

\section{Origin of the $1/f$ fluctuations in the power spectrum}
\label{sec:origin}
In this section, we investigate the origin of the $1/f$ fluctuations
in the power spectrum $P_z(\omega)$
that we found out in the previous section.
The relations between the physical quantities and figures are summarized
in Table~\ref{tab:summary-of-figures}.
To begin with, we look more closely at the time evolution of the
orbits which were analyzed in the previous section (see Fig.~\ref{fig:ps}).

First, in Fig.~\ref{fig:poin13}, we plot the time series of the $v^r$ components in the
Poincar\'{e} map with $z=0$ and $v^{\theta}<0$.
We choose parameter sets in Figs.~\ref{fig:poin13} (a) and \ref{fig:poin13} (b)
with the same values as those in Figs.~\ref{fig:pmap} (a) and
\ref{fig:pmap} (b), 
respectively. 
In Fig.~\ref{fig:poin13} (a) the values of the $v^r$ components 
in the Poincar\'{e} map fluctuate with stagnation.
It seems that the duration of the stagnation varies from time to time, 
and is a mixture of various lengths. There is  no typical one.
This feature is observed commonly when the power spectrum $P_z(\omega)$
is the $1/f$ type.
Then we expect that this behavior is closely related to the $1/f$ fluctuations
in the power spectrum $P_z(\omega)$ in Fig.~\ref{fig:ps} (a).
In Fig.~\ref{fig:poin13} (b), on the other hand, 
the values of the $v^r$ components in the Poincar\'{e} map
oscillate almost monotonously.
This behavior is consistent with the power spectrum $P_z(\omega)$
showing white noise in Fig.~\ref{fig:ps}(b).

Second, in Fig.~\ref{fig:orb},
we plot the same orbit we shown in Fig.~\ref{fig:poin13} (a)
in the two-dimensional configuration space $(r,z)$.
We find that the whole orbit (Fig.~\ref{fig:orb} (a))
has two significant components (Figs.~\ref{fig:orb} (b) and
\ref{fig:orb} (c)).
The periods when the orbit stagnates around each component in Fig.~\ref{fig:orb}
perfectly correspond to that when
$v^r$ components in the Poincar\'{e} map stagnate in
Fig.~\ref{fig:poin13} (a).
We never see such stagnant motions when the power spectrum become white noise
(Fig.~\ref{fig:poin13}(b)).
These results suggest that the $1/f$ fluctuations 
in the power spectrum $P_z(\omega)$ are originated by
the stagnation and the stickiness around these two distinguishable components,
while traveling back and forth between them, in the particle motions.

Now,
let us show a mechanism by which the motion is stagnant or not, as shown in
Figs.~\ref{fig:poin13} and \ref{fig:orb};
that is,
the power spectrum $P_z(\omega)$ shows $1/f$ or white noise as shown in Figs~\ref{fig:ps},
depending on the combination of $(J,S)$, as shown in Fig.~\ref{fig:pdia}.
The essence of the mechanism is that 
such stagnant motions are originated around the tori (regular orbits), 
and the structure of the tori changes
depending on the combination of $(J,S)$.
To see the structure of tori more clearly, it is useful to see the motions at lower energy where 
the chaotic sea shrinks and tori get larger.
In Fig.~\ref{fig:pmap-low} we plot the Poincar\'{e} maps with 
same values of $J$ and $S$ as in Fig.~\ref{fig:pmap} but smaller values
of $E$. 
Around this value of energy $E$, the structure of tori is not very
sensitive to the change of $E$, compared to the changes of $J$ or $S$.
The volume of the chaotic sea in the Poincar\'{e} maps in
Fig.~\ref{fig:pmap-low} is smaller than that in Fig.~\ref{fig:pmap}.
However, we will see that the topological structure of the phase space
is quite similar to that of Fig.~\ref{fig:pmap}.

First, we pay attention to the chaotic orbits which we saw in
Fig.~\ref{fig:pmap-low}.
In Fig.~\ref{fig:poin13-low}
we plot the time series of the $v^r$ components in the Poincar\'{e} maps 
Fig.~\ref{fig:pmap-low}.
We adopt each data in Figs.~\ref{fig:poin13-low} (a) and
\ref{fig:poin13-low} (b)
as the chaotic orbits that are plotted
in Figs.~\ref{fig:pmap-low} (a) and \ref{fig:pmap-low} (b),
respectively.
It is found that the values of the $v^r$ components
fluctuate with stagnation in Fig.~\ref{fig:poin13-low} (a),
while they do not in Fig.~\ref{fig:poin13-low} (b).
The feature about the stagnation in Figs.~\ref{fig:pmap-low} (a) and
\ref{fig:pmap-low} (b) is similar to that in Figs.~\ref{fig:poin13} (a)
and \ref{fig:poin13} (b), respectively.
In Fig.~\ref{fig:orb-low}, we plot the same orbit we showed in Fig.~\ref{fig:poin13-low} (a)
in the two-dimensional configuration space $(r,z)$.
Similarly to the higher-energy case in Figs.~\ref{fig:poin13} and \ref{fig:orb},
the periods when the orbit stagnates around each component in Fig.~\ref{fig:orb-low}
perfectly correspond to that when
$v^r$ components in the Poincar\'{e} map stagnate in
Fig.~\ref{fig:poin13-low}(a).
Now, in Fig.~\ref{fig:ps-low}
we plot the power spectrum $P_z(\omega)$ of the same chaotic orbits as
Figs.~\ref{fig:poin13-low} (a) and \ref{fig:poin13-low} (b).
It is clearly confirmed that the pattern of the power spectrum $P_z(\omega)$
shows $1/f$ in
Fig.~\ref{fig:ps-low} (a),
while the pattern shows white noise in
Fig.~\ref{fig:ps-low} (b).
Therefore, similarly to the higher-energy case, it is suggested that 
the $1/f$ fluctuations in the power spectrum $P_z(\omega)$ are
originated by the motion which 
stagnates and sticks around the two distinguishable
components (Figs.~\ref{fig:orb-low} (b) and \ref{fig:orb-low} (c)),
while traveling back and forth between them.
Our results
indicate that the pattern of the power spectrum $P_z(\omega)$ of the chaotic orbits
strongly depends on the value of the total angular momentum $J$ and spin $S$,
while the value of energy $E$ does not affect the pattern of the power
spectrum so strongly, although it affects the volume of the tori.

Second, we pay attention to the tori (regular orbits) which we saw
in Fig.~\ref{fig:pmap-low}.
In Fig.~\ref{fig:torus-low} we plot the orbits of the tori 
in the two-dimensional configuration space $(r,z)$.
Each color of lines in Figs.~\ref{fig:torus-low} (a) and \ref{fig:torus-low} (b) 
corresponds to that of closed curves in Figs.~\ref{fig:pmap-low} (a) and
\ref{fig:pmap-low} (b). 
For example, the orbits A and B in Fig.~\ref{fig:torus-low} (a)
correspond to the closed curves A and B in Fig.~\ref{fig:pmap-low} (a).
Indeed, the existence of some regular orbits in this model
have been shown in the paper~\cite{sm97}.
In this paper, however, we should emphasize
that the structure of the phase space
is quite different between panels
(a) and (b) in Fig.~\ref{fig:pmap-low} or Fig.~\ref{fig:torus-low}.
For example, any tori which correspond to A and B 
in Fig.~\ref{fig:pmap-low} (a) or Fig.~\ref{fig:torus-low} (a)
can never be found in Fig.~\ref{fig:pmap-low} (b) or Fig.~\ref{fig:torus-low} (b).
That is, we should emphasize that the topological structure of the tori changes
 depending on the parameter sets $(J,S)$, and the change is in accordance with
 the change of types of the power spectra 
(see Figs.~\ref{fig:ps-low} and \ref{fig:torus-low}).
Moreover, it is quite important to note that the orbits of the tori A and B in
Fig.~\ref{fig:torus-low} (a) resemble the chaotic orbits in
Fig.~\ref{fig:orb} (c) and Fig.~\ref{fig:orb} (b), 
or Fig.~\ref{fig:orb-low} (b) and Fig.~\ref{fig:orb-low} (c), 
respectively.
This resemblance means that the orbits in Figs.~\ref{fig:orb} and
 \ref{fig:orb-low}
stagnates around the tori A and B in Fig.~\ref{fig:torus-low} (a),
although it is not easy to detect such tori in Fig.~\ref{fig:pmap} (a)
since chaotic sea dominates and the tori A and B shrink extremely
with the larger value of the energy $E$.

Summing up the above results leads us to the following conclusions.
The $1/f$ fluctuations 
in the power spectrum are originated by
the orbit which stagnates around the tori while traveling 
back and forth between them.
Whether we can observe $1/f$ fluctuations
in the power spectrum in the time series of a component
depends on the topological structure of the phase space.
It is the pair of tori A and B in Figs~\ref{fig:pmap-low} (a) and \ref{fig:torus-low} (a)
that gives rise to the $1/f$ fluctuations in the power spectrum $P_z(\omega)$.
Moreover, our results suggest that the topological structure of the tori
changes strongly depending on 
the total angular momentum $J$ and spin $S$ rather than the energy
$E$. Therefore, it is reasonable that 
the pattern of power spectrum $P_z(\omega)$ of the
chaotic motion depends mainly on the
parameter sets $(J,S)$, and we can expect that the phase diagram that we obtained in the
previous section (see Fig.~\ref{fig:pdia}) is reliable.

\section{Summary and Discussion}
\label{sec:summary}
In this paper we have characterized the properties of chaos in a spinning 
test particle in Schwarzschild spacetime.
We have calculated the power spectrum of the time series of $z$ 
components of the test particle's position, $P_z(\omega)$,
and  found out that the pattern of the power spectrum $P_z(\omega)$ is $1/f$
or white noise in the low-frequency range (see Fig.~\ref{fig:ps}).
That is, 
we have succeeded in classifying the chaotic motions, which had been
classified as merely chaotic in the paper~\cite{sm97}, into these two
distinct types (see Table~\ref{tab:classification}).
The important point is that the pattern of the power spectrum
strongly depends on the spin $S$ and 
the total angular momentum $J$ of the test particle 
and not on the initial conditions.
Our analyses also suggest that the value of the energy $E$
does not affect so strongly whether the power spectrum 
$P_z(\omega)$ becomes $1/f$-type spectrum or not 
(see Figs.~\ref{fig:ps} and \ref{fig:ps-low}).
Then, testing the pattern of $P_z(\omega)$ for various
grid points in two-dimensional $(J,S)$ plane,
we have obtained the phase diagram for 
the character of the chaotic motions
(see Fig.~\ref{fig:pdia}).
This phase diagram enables us in principle
to guess the properties of the system 
($J$ and $S$) by observing the dynamics of the test particle, 
even if the motion is chaotic.


Furthermore we have pointed out that the pair of tori A and B in
Figs~\ref{fig:pmap-low} (a) and \ref{fig:torus-low} (a)
gives rise to the $1/f$-type power spectrum of the time series of $z$
components of the
particle's position $P_z(\omega)$.
The important point is that the chaotic orbits stick near the tori,
while traveling back and forth between them.
Whenever the power spectrum $P_z(\omega)$ becomes the $1/f$
spectral pattern,
we have found that the orbit (Fig.~\ref{fig:orb} (a))
stagnates around two significant components (Figs.~\ref{fig:orb} (b) and
\ref{fig:orb} (c)).
Moreover, investigating the motions with the value of lower energy where more
tori dominate,
we have confirmed that such significant components imitate the orbits of the tori
characterized as $A$ and $B$ in Figs.~\ref{fig:pmap-low} (a) and
\ref{fig:torus-low} (a).

Eventually, the conclusion is summarized as follows.
We have two types of chaos as we have seen by the power
spectra in the system we have studied in this paper.
One is $1/f$, and the other is white noise.
The difference of the properties of chaos or the power spectra is
caused by the topological structure of the phase space,
which in turn is governed by the physical parameter set $(J,S)$ of the system.
From this point of view, the chaos we found in this paper is not always
merely random.

The type of motion where the phase point in chaotic orbit stays close
to some regular orbits (tori) for some long time is known as
``stagnant motion'' or  ``sticky motion,''
 and is often observed in Hamiltonian dynamical
systems~\cite{ka83,ch84,ai89}.  Stagnant motions are usually accompanied by
$1/f$ fluctuations and are considered to be due to
the fractal structure of the phase space~\cite{ai84,me85,yy98}.
In particular, stagnant motions are often observed for weakly chaotic,
nearly integrable systems.
This is consistent with the fact that we have observed the $1/f$ fluctuations 
for smaller values of spin $S$ when the chaos is weak,
and white noise for larger values of spin $S$ when the chaos is strong.
Until now such $1/f$ fluctuations have not been discovered in any relativistic systems. 
We have shown that the $1/f$ fluctuations 
we observed for the first time in the relativistic system are also
generated by such stagnant motion,
that the particle motion stagnates around regular orbits, 
while traveling back and forth between them
in Schwarzschild spacetime.


Finally the astrophysical implications of our results should be mentioned.
It is true that the value of the spin $S$
where the particle motion is remarkably chaotic
is so large that such a star cannot exist,
which has been remarked in the paper~\cite{sm97}.
Then the result obtained in this paper is relevant as an illustration and is
not directly applicable to a physical system.
However, it is expected that the method and theory in this paper
can be useful and applied to other astrophysical systems in practice,
since nature is filled with phenomena that exhibit chaotic behavior.

\acknowledgements
We would like to thank Kei-ichi Maeda for valuable discussions.
H.K. and K.K. are supported by JSPS Fellowship for Young Scientists.
The authors are grateful to the anonymous referee for helpful comments
on improving the text and the figures.


\newpage
\begin{table}[htbp]
  \centering
  \caption{summary of figures}
  \label{tab:summary-of-figures}
  \begin{tabular}{|c||c|c|}\hline
 $S$ & $1.2\mu M$ & $1.4\mu M$ \\ \hline
$P_z(\omega)$ & $1/f^\nu$ (Figs.~\ref{fig:ps} (a), \ref{fig:ps-low} (a))
   & white noise (Figs.~\ref{fig:ps} (b), \ref{fig:ps-low} (b))\\ \hline
$v^r(t_i)$ & 
stagnating (Figs.~\ref{fig:poin13} (a), \ref{fig:poin13-low} (a)) &
 non stagnating (Figs.~\ref{fig:poin13} (b), \ref{fig:poin13-low}
   (b))\\ \hline
Tori A, B &
 observed (Figs.~\ref{fig:orb}, \ref{fig:pmap-low} (a),
   \ref{fig:orb-low}, \ref{fig:torus-low} (a)) & not observed
   (Figs. \ref{fig:pmap-low} (b), \ref{fig:torus-low} (b))\\ \hline 
  \end{tabular}
\end{table}

\begin{table}[htbp]
\caption{Schematic classification of motions in this system}
\label{tab:classification}
\[
\begin{cases}
\text{bounded} & \cdots 
\begin{cases}
\text{regular} \ & \ \cdots \ \text{(periodic or quasi periodic)}
\\
& \\
\text{chaotic} \  & \ \cdots \  \begin{cases}
1/f^\nu \ & \ \cdots \   \text{(correlated)}  \\
\text{white noise} \ & \ \cdots \ \text{(uncorrelated)}
\end{cases}
\end{cases}\\
\text{unbounded} 
\end{cases}
\]
\end{table}
\newpage
\begin{figure}
\caption{The schematic drawing of the Poincar\'{e} maps.} 
\label{fig:poin}
\end{figure}
\begin{figure}
\caption{The Poincar\'{e} maps with $z=0$ and $v^{\theta}<0$. All orbits
have the total angular momentum $J=4.0\mu M$. The magnitude of 
spin and the total energy are $S=1.2\mu M$ and $E=0.93545565\mu$ in
 panel (a), and $S=1.4\mu M$ and $E=0.92292941\mu$ in
 panel (b), respectively.
The dots with different colors correspond to the data from the orbits 
with different initial conditions.}
\label{fig:pmap}
\end{figure}
\begin{figure}
\caption{The power spectrum of the time series of $z$ 
components of the particle. Each set of the parameters and the initial conditions 
in panels (a) and (b) are the same as in Figs.~\ref{fig:pmap} (a) and (b),
 respectively.
The long-time correlations with the power law, so called $1/f$ fluctuations, 
are observed in panel (a), while no such correlations are
 observed in panel (b).}
\label{fig:ps}
\end{figure}
\begin{figure}
\caption{The phase diagram for the type of the power spectrum pattern.
The patterns of the power spectrum for the chaotic orbits at 
grid points in a two-dimensional $(J,S)$ configuration are tested. 
At the points where circle symbols ($\bigcirc$) are marked,
the $1/f$-type power spectrum is observed.
At the points where cross symbols ($\times$) are marked,
the white-noise power spectrum is observed.
At the points where triangular symbols ($\triangle$) are marked,
the orbit apparently behaves 
almost regular in the temporal interval of numerical computation.
In the blank region,
the test particle goes to infinity or falls into the black hole,
since the energy surface is unbounded.
}
\label{fig:pdia}
\end{figure}
\begin{figure}
\caption{The time series of the $v^r(t)$ components of the Poincar\'{e} map with
$z=0$ and $v^{\theta}<0$.
Parameter sets $(J,S,E)$ in panels (a) and (b) are chosen so as to
 have the same values as those
in Figs.~\ref{fig:pmap} (a) and (b), respectively.}
\label{fig:poin13}
\end{figure}
\begin{figure}
\caption{The orbit which we saw 
in Fig.~\ref{fig:poin13} (a) in the two-dimensional configuration space $(r,z)$.
(a) The orbit for the whole period $0<t<50000$. (b) The orbit for the period $4000<t<11000$.
(c) The orbit for the period $24000<t<28000$.}
\label{fig:orb}
\end{figure}

\begin{figure}
\caption{The Poincar\'{e} maps with $z=0$ and $v^{\theta}<0$.
Parameter sets $(J,S)$ in panels (a) and (b) are chosen as the same values as those
in Figs.~\ref{fig:pmap} (a) and \ref{fig:pmap} (b), respectively.
The values of the total energy are chosen as $E=0.933\mu$ (panel (a)) and
 $E=0.9205\mu$ (panel (b)), that are smaller
than that of Fig.~\ref{fig:poin13} (a) and \ref{fig:poin13} (b), respectively.}
\label{fig:pmap-low}
\end{figure}

\begin{figure}
\caption{The time series of the $v^r(t)$ components of the Poincar\'{e} map with
$z=0$ and $v^{\theta}<0$ .
Each orbit corresponds to that of the chaotic orbits in
Figs.~\ref{fig:pmap-low} (a) and \ref{fig:pmap-low} (b), respectively.
}
\label{fig:poin13-low}
\end{figure}

\begin{figure}
\caption{The orbit which we saw in Fig.~\ref{fig:poin13-low} (a) in the
 two-dimensional configuration space $(r,z)$.
(a) The orbit for the whole period $0<t<70000$. (b) The orbit for the period $22000<t<38000$.
(c) The orbit for the period $42000<t<70000$.}
\label{fig:orb-low}
\end{figure}

\begin{figure}
\caption{The power spectrum $P_z(\omega)$ of
the chaotic orbits in Figs~\ref{fig:pmap-low} (a) and \ref{fig:pmap-low} (b).
We adopt the time-series $z(t)$ in panels (a) and (b) as
the chaotic orbits in
Figs.~\ref{fig:pmap-low} (a) and \ref{fig:pmap-low} (b), respectively.
The long-time correlations with power law, so called $1/f$ fluctuations, 
are observed in panel (a), while no such correlations are
 observed in panel (b).}
\label{fig:ps-low}
\end{figure}

\begin{figure}
\caption{ The orbits in the two-dimensional configuration space $(r,z)$
corresponding to each torus in Figs.~\ref{fig:pmap-low} (a) and
 \ref{fig:pmap-low} (b).
Each line color in panels (a) and (b) corresponds to that of the
closed curves in Figs.~\ref{fig:pmap-low} (a) and \ref{fig:pmap-low} (b), respectively. 
For example, the regular orbits A and B in Fig.~\ref{fig:orb-low} (a)
correspond to the tori A and B in Fig.~\ref{fig:pmap-low} (a).
}
\label{fig:torus-low}
\end{figure}

\end{document}